\documentclass[pre,twocolumn,showpacs]{revtex4}
\usepackage{amssymb}
\usepackage{graphicx}
\usepackage{amsmath}

\begin{document}

\input{epsf}

\title{Effect of point defects on heat capacity of 
 yttria-stabilized zirconia}
\author{S.~Ostanin$^{1}$ and E.~Salamatov$^2$}
\address {$^1$Department of Physics, University of Warwick, Coventry CV4 7AL, UK}
\address {$^2$Physico-Technical Institute, Ural Branch of RAS, 
 132 Kirov Str., 426001 Izhevsk, Russia}

\begin{abstract}
 First-principles calculation and anharmonic dynamical theory 
 were used in sequence to explain a large excess heat capacity
 observed in yttria-stabilized zirconia 
 in comparison with the additive rule value.
 It is found that the excessive shape of heat capacity decays 
 gradually with the Y$_{2}$O$_{3}$ doping when the number of 
 environmentally different O sites falls to its zero value 
 at 33 mol $\%$ Y$_{2}$O$_{3}$-ZrO$_2$ due to the Y atoms adjacent.
 The model and results presented in this work provide a key insight
 into the complex behaviour and characterization of fast-ion conductors.      
\end{abstract} 

\maketitle


 Point lattice defects both native and artificial, 
 such as vacancies, self-interstitials, solute and substitute 
 atoms, strongly affect the most fundamental properties of 
 materials.\cite{Flynn} In yttria-stabilized zirconia (YSZ), 
 the material with numerous commercial applications
 \cite{Subbarao}, formed by the addition of Y$_{2}$O$_{3}$ to 
 ZrO$_{2}$, the trivalent dopant Y$^{3+}$ substitute for some of
 the host cations and, in order to maintain charge neutrality, one O
 vacancy (V$_0$) must be created for each pair of dopant cations.
 Y$_{2}$O$_{3}$ is used to stabilize the tetragonal ($t$) phase of
 (ZrO$_{2}$)$_{100-x}$(Y$_{2}$O$_{3}$)$_{x}$ over the composition
 range 2$< x <$9 mol $\%$ and the cubic ($c$) fluorite phase with
 4$<x<$40 mol $\%$. 
 The presence of relaxed V$_0$ and Y substitute atoms 
 make the local atomic environments of YSZ rather different 
 from the stoichiometric high-temperature $t$- and 
 $c$-polymorphs of pure ZrO$_{2}$ whose 8-fold cations are distorted 
 and perfectly coordinated, respectively.
 In YSZ, the average cation coordination number, ranged
 between 7 and 8, is reduced gradually with increasing Y$_2$O$_3$.
 If the V$_0$ associates with Zr ions, as the 
 X-ray absorption findings \cite{Li} suggest, it may support a 
 coordination-driven ordering model of stabilization.\cite{Stefanovich} 
 Placing Y in the next nearest neighbour (NNN) cation positions allows 
 the coordination of Zr in the nearest neighbours (NN) sites 
 to the V$_0$ to be similar to the monoclinic ZrO$_{2}$ arrangement, 
 with Zr located in 7-fold coordination environments, 
 while Y remains 8-fold coordinated. 

 The electron energy-loss near-edge structure 
 (ELNES) of YSZ demonstrates the features of the
 experimental O $K$-edge shapes which depend on the crystal structures
 and the Y$_{2}$O$_{3}$ composition.\cite{OCMVAFP}
 In modelling the most important and widely used properties of YSZ the
 relaxation mechanism is essential. The effects of relaxation were 
 treated within the plane-wave, pseudopotential based free energy 
 molecular dymanics (FEMD) technique.\cite{Ali} 
 Using the relaxed configurations (RC) of point defects, the ELNES 
 calculations were carried out.\cite{OCMVAFP}
 Since the results of calculation show very good agreement with the 
 experimental O $K$-edge signal, it seems that theory reflects the 
 realities of relaxation. There is much yet to be learned 
 about the fundamental nature of transport processes in YSZ. Thus, our 
 motivation is to make a contribution to the microscopic modelling 
 and tailoring the heat capacity of YSZ. 

 A large excess heat capacity (EHC) has been observed \cite{heat_cap} 
 in YSZ at low and room temperatures for 7.8, 9.7 and 11.4 mol $\%$ 
 Y$_{2}$O$_{3}$ compared to the heat capacity calculated from those of 
 pure ZrO$_2$ and Y$_{2}$O$_{3}$ by the additivity rule:
 ${\Delta}C_p$ = ${C}_{p}^{YSZ}$ - [(1-$x$) ${C}_{p}^{ZrO_2}$ + 
 $x$ ${C}_{p}^{Y_{2}O_{3}}$]. The shape of ${\Delta}C_p$($T$) 
 is very similar to that of the Schottky disorder. 
 The same behaviour of heat capacity may yield the two-level 
 system (TLS), quite a simple model, used to simulate the 
 low-temperature properties of amorphous insulators:
 $C_p^{TLS}$=$k_B (\Delta /T)^2$ $e^{-\Delta /T}$/($e^{-\Delta 
 /T}$+1)$^2$, where $\Delta$ is the energy difference between the levels. 
 $C_p^{TLS}$ shows a maximum at temperature $T_m \simeq$ 0.42~$\Delta$ 
 when the filling-rate variation for the occupied TLS states 
 reaches its maximum value. At $T \gg T_m$, all TLS states 
 are occupied equally and therefore heat capacity
 decreases as $C_p^{TLS} \sim 1/{T^2}$.

 
 It appears that for an anharmonic oscillator (AO),
 which dynamics is defined by the double-well
 potential $U$($x$)=$-x^2$+$a x^3$+$b x^4$, ($a,b >$0) with
 a maximum at $x_0$=0 and two minima at $x_1$ and $x_2$,
 separated by the energy $\Delta$=$U(x_2)-U(x_1) >$0, 
 the temperature dependence of heat capacity is similar 
 to that of TLS. The partition function 
 $Z_{AO}$=$\frac{1}{2 \pi h}$
 $\int dx dv \, exp[-\frac{m v^2/2 + U(x)}{k_B T}]$  
 is calculated assuming $\mid U_0 \mid \gg \mid U_i \mid$.
 If $U$ is approximated near its minima 
 by the second-degree polynomials, 
 $U_1 \simeq (x-x_1)^2 {\omega}_1^2 /2$ and
 $U_2 \simeq (x-x_1)^2 {\omega}_2^2 /2 + \Delta$
 where $\omega _i$ are the frequencies of small vibrations at $x_i$
 then, using the general thermodynamic relations, one can obtain
 $C_p^{AO}$ = $k_B$\,$\frac{\Delta^2}{T^2}$\,$\frac{exp(-\Delta/T)}
 {exp(-\Delta/T + \omega_1 /\omega_2)^2}$ + $k_B$. 
 In the latter, the temperature dependence of the first term, 
 $\Delta C$($T$), which defines the difference in heat capacity 
 between anharmonic and harmonic oscillators,
 is close to $C_p^{TLM}$ and finally, at $\omega_1 /\omega_2$=1,
 $\Delta C \equiv C_p^{TLM}$. Hence, the EHC effect may 
 appear as the result of strong anharmonic vibration modes
 either localized or delocalized ones. 

 The fact that at $x >$8 mol $\%$ Y$_2$O$_3$ ${\Delta}C_p$ 
 decreases \cite{heat_cap} with increasing the dopant concentration
 may suggest that the delocalized anharmonic vibrations
 cause the EHC effect. It is known that the main contribution to the 
 structural transformations comes from particular vibrational modes.
 In pure ZrO$_{2}$, a zone-boundary soft phonon, $X_{2}^{-}$, 
 which breaks the $c$-symmetry of the O sublattice, displacing
 the O atoms toward their positions in the $t$-phase, may be 
 responsible for the $c-t$ phase transformation.
 In YSZ, the experimental data \cite{Exp-neutron} are not fully 
 clarified for this soft mode because of the static disorder 
 in the O sublattice. Using the RC of YSZ, the $X_{2}^{-}$-like 
 phonon was calculated \cite{OS2002} 
 by means of the frozen-phonon method for each composition of 
 (ZrO$_{2}$)$_{100-x}$(Y$_2$O$_3$)$_x$ between 3 and 10 
 mol $\%$ Y$_2$O$_3$. 
 For 10 mol $\%$ Y$_{2}$O$_{3}$, the effective potential
 has a single minimum and, with decreasing Y$_{2}$O$_{3}$ content, 
 the potential develops two minima. 
 The temperature dependence of the phonon frequency, calculated 
 within the modified pseudoharmonic approximation,
 quantifies accurately the transition temperature above which the 
 $c$-phase is stabilized.\cite{OS2002}
 In the 3 mol $\%$ Y$_2$O$_3$ case, it may start around room temperature. 
 The upper boundary of the single $t$-phase field in the experimental 
 phase diagram \cite{Miller81} shows a similar rapid drop of temperature 
 with Y$_2$O$_3$ concentration. 
 However, the $X_{2}^{-}$-like soft phonon of YSZ, in another words,
 the delocalized vibrations can not be responsible for the EHC effect 
 because of the potential symmetry. 

 Regarding anharmonic effects, the atomic vibrations localized near 
 the V$_0$ should be considered as well. YSZ is a solid
 electrolyte in which ionic transport takes place by anions 
 moving among their positions by the vacancy diffusion mechanism. 
 If any two V$_0$ sites, which an O atom occupies before and after 
 a diffusive jump, are non-equivalent then theory enables the 
 double-well asymmetric potential profile.
 In such a study of diffusion, the molecular dynamics technique can 
 fruitfully be used. The principal output of the dynamic 
 computation is the set
 of atomic trajectories from which all other properties are calculated. 
 Molecular dynamics has been used in the past
 by many authors to calculate the potential energy dependant on the 
 position of diffusion atom.\cite{Flynn} Even if one could calculate
 the induced trajectories through the saddle-point 
 the results were strongly affected by the reaction coordinates
 of all other atoms. The benefits and limitations 
 of this molecular dynamics approach have been discussed and it is 
 generally accepted now that using the large unit cells this scheme is 
 very complicated for precise calculation whereas the results, obtained 
 in such a way, are not always satisfactory.\cite{Bennett}

 The features of the computed $U$ profile were demonstrated
 using the single-$V_0$ 11-atom Zr$_2$Y$_2$O$_7$ cell, which contains two 
 formula units of ZrO$_2$ and one formula unit of Y$_2$O$_3$, 
 that corresponds to a composition of 33 mol $\%$ 
 Y$_2$O$_3$.\cite{OCMc+2000} 
 The RC, shown in the (a) panel of Fig.~1, was used in a further 
 static optimization using the displacements of O-2 along 
 the $z$-direction while all other atoms are fixed at their relaxed 
 positions. In this case, labeled in the panel (b) of Fig.~1 as 
 a ``solid pass'' trajectory, the O-2 position in the $z$=0 plane 
 corresponds to the saddle point with the potential barrier $E_b$ 
 of $\sim$0.2~eV/cell. 
 If Zr in the $z$=0 plane were allowed to relax when the O-2 is moved,
 as shown in Fig.~1 as the ``soft pass'' trajectory, then the equlibrium 
 position of O-2 is much closer to the saddle point  while $E_b$ becomes 
 significantly less.
\begin{figure}
\resizebox{0.9\columnwidth}{!}{\includegraphics*{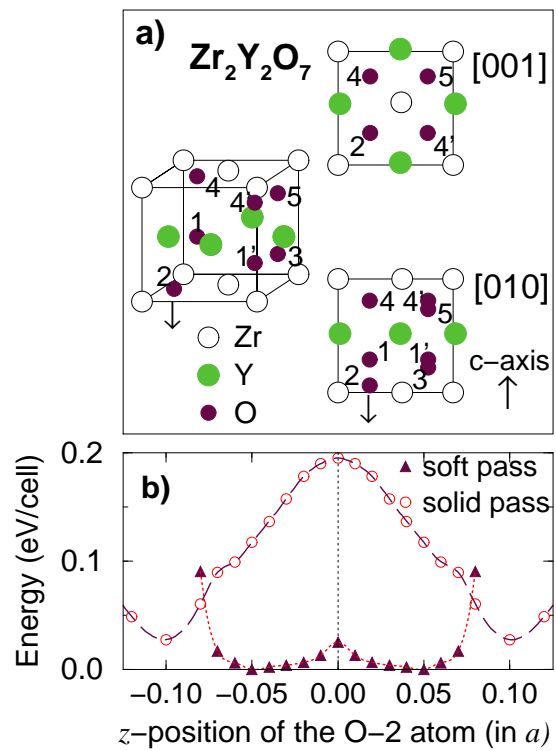}}
\caption{Model of Zr$_{2}$Y$_{2}$O$_{7}$ after relaxation
 is shown in the (a) panel. The plan views shown in (a) represent
 the views along the [001] and [010] directions. The O-2 displacement
 along [001] is marked by an arrow. In the (b) panel, the curve, 
 labelled as solid pass, shows the change in the FEMD energy
 as a single O-2 atom is moved from its equilibrium position
 along the $z$ axis while all other atoms are fixed at their
 relaxed positions. The soft-pass curve, shown after shifting to
 a common zero energy, corresponds to the O-2 displacements when its
 nearest Zr may relax.}
\label{fig:MODEL}
\end{figure} 
 In Zr$_2$Y$_2$O$_7$, where all O sites are equivalent to each other, 
 the Zr/Y environment enables the symmetric form of $U(x)$ and 
 therefore can not result in the EHC effect. The saddle-point 
 neighorhood plays a key role on the instability-barrier formation, 
 which is very sensitive to the model parameters.
 Thus, the dynamics theory of ion transport is still very much 
 a semiempirical science and further simplification,
 based on the realities for the activation energy, should be made.

 Atomistic RC of the (96-$y$)-atom supercells ($y$=1,2,...) allow to 
 model (ZrO$_{2}$)$_{100-x}$(Y$_2$O$_3$)$_x$. In the 95-atom cell, which 
 corresponds to $\sim$3 mol $\%$ Y$_{2}$O$_{3}$, the  
 RC with both Y atoms in the NNN shell was obtained.\cite{OCMVAFP} 
 Fig.~2 shows the RC energies,
 expressed as an equivalent temperature per 12-atom cell,
 to give an idea of the relative stability and temperature required.
 In the 94-atom cell, the RC with two V$_0$s 
 along the $\langle$111$\rangle$ fluorite direction results in 
 a 6-fold coordinated and six 7-fold coordinated NN Zr atoms.
 This result supports the interpretation of the neutron data by
 Goff et al. \cite{Goff99}.
\begin{figure}
\resizebox{0.9\columnwidth}{!}{\includegraphics*{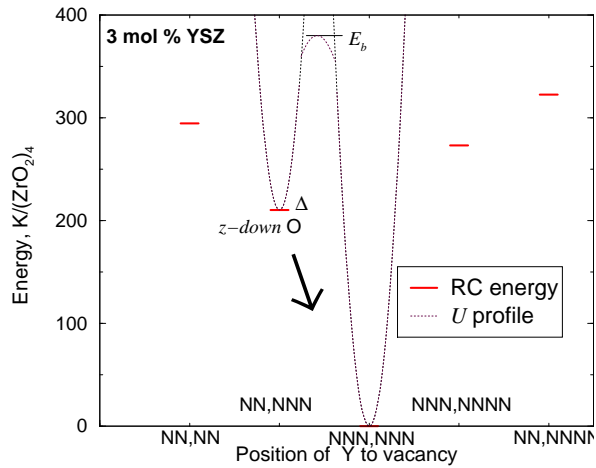}}
\caption{Relaxed energies of the 95-atom
 YSZ cell are plotted vs positions of the two Y dopant cations
 on the NN or/and NNN cation shells. The diffusion jump of the
 lower apical NN O allows to construct the double-levels potential
 $U$ shown in the diagram as a sketch.}
\label{fig:RELAXATION}
\end{figure} 
 Electrostatic considerations suggest that V$_0$s should repel.
 From this point of view, the tri-V$_0$ RC,
 with all V$_0$ separated by $\sqrt{5}$/2 from each other (in units 
 of the $c$-ZrO$_2$ lattice constant $a$), is rather reasonable.
 The repulsion tendencies between vacancies consistent with both 
 electrostatics and the Zr-coordination lowering model. Since
 the shortest distance of $a$/2 between the V$_0$ sites is not 
 energetically favorable in YSZ at low Y$_2$O$_3$ concentrations 
 we believe that each V$_0$ has six NN O atoms each of
 which can jump into the empty site.  

 The intrinsic process of such chaotic O moves should occur in YSZ at 
 finite temperatures via the vacancy mechanism. 
 In particular, when an apical O in 3 mol $\%$ Y$_{2}$O$_{3}$
 moves into V$_0$ the RC with the Y NN and NNN atoms is created. 
 The difference $\Delta $=210~K between these two RC was used to 
 construct the potential profile, as shown in the lower diagram of Fig.~2.   
 If V$_0$ moves consequently into the next O site of this anion tunnel
 along the $z$ axis, the Y NN-NNNN configuration is realized. 
 We take $E_b = E_a \simeq$1~eV using the  
 activation energy $E_a$ of YSZ, obtained from the Arrhenius relation
 for the ionic dc conductivity \cite{Arrhenius}
 $\sigma \sim \omega_0 d^2 exp(-E_a/k_B T)$, where $\omega_0$
 is the attempt frequency. 
 To form the effective potential, characterized by $E_b$, $\Delta$ and
 the distance moved during a single jump $d$=$\mid x_1-x_2 \mid$, 
 the three-parameter polynomial $U(x)$= $-a x^2 + b x^3 + c x^4$ was used.
 Since $E_b \gg \Delta$, the O vibrations near the $U$ minima may be 
 estimated as $\omega \approx \sqrt{ \frac {32 E_b} {m d^2} }$, where 
 $m$ is the O mass. As a result, $\omega \approx$ 5$\times$10$^{14}$Hz 
 is close to that found experimentally \cite{Ohta+2002} 
 showing very reasonable model approximation.  


 In Fig.~3a the calculated EHC shape 
 $\Delta C_p$($T$) = ${\mathcal N}^{TLS} \cdot C_p^{TLS}$ 
 is plotted, with its dependence on temperature and dopant concentration. 
 Here ${\mathcal N}^{TLS}$ was obtained in such a way that each maximum
 coincides with the experimental EHC amount \cite{heat_cap} at 7.76, 9.7 
 and 11.3 mol $\%$ Y$_{2}$O$_{3}$. The EHC observations of 7.76 
 mol $\%$ Y$_{2}$O$_{3}$ are shown in Fig.~3a to illustrate the similarity. 
 All theoretical $\Delta C_p$ shapes, with a maximum at 88~K,
 have been aligned with the experimental peak at $T_m$=75~K using
 the $T/\Delta$ dimensionless scale.
\begin{figure}
\resizebox{0.9\columnwidth}{!}{\includegraphics*{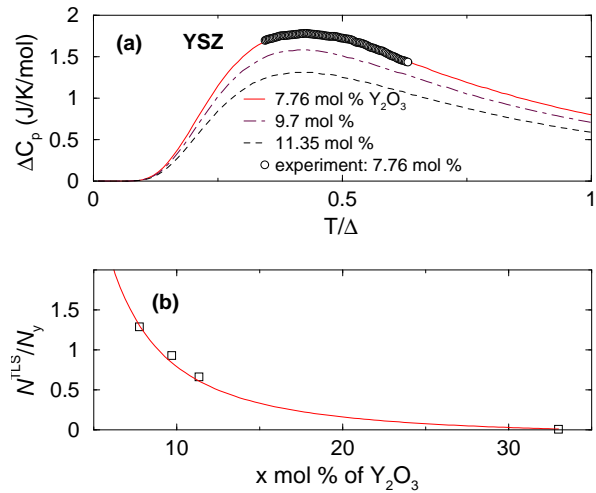}}
\caption{Calculated $\Delta C_p$ for the 7.76, 9.7 and 11.35 mol  
 $\%$ Y$_2$O$_3$-ZrO$_2$ are shown vs $T/\Delta$
 in the (a) panel compared with the experimental EHC shape
 of 7.76 mol $\%$ Y$_2$O$_3$.
 In the panel (b), the ratio of ${\mathcal N}^{TLS}$,
 obtained using $\Delta C_p$, to the number of vacancies
 ${\mathcal N}_y$ are shown as $\Box $ for each Y$_2$O$_3$
 concentration and connected by eye.}
\label{fig:CAPACITY}
\end{figure}
 The (b) panel of Fig.~3 shows that the
 ${\mathcal N}^{TLS}$/${\mathcal N}_y$ ratio (${\mathcal N}_y$ is
 the number of $V_0$) falls quickly with increase in Y$_2$O$_3$ content
 and, finally, 
 ${\mathcal N}^{TLS}$/${\mathcal N}_y \to$0 at $x$=33 mol $\%$ when the 
 $U$ symmetry for diffusion atom should appear. 
 At low concentrations of Y$_{2}$O$_{3}$, not all  
 possible diffusion jumps lead to the asymmetric $U$. 
 For example, in 3 mol $\%$ Y$_{2}$O$_{3}$, ${\mathcal N}_y$=1 whereas 
 ${\mathcal N}^{TLS}$ falls from six to four since two O NN sites 
 to V$_0$ are energetically equivalent due to the Y atoms adjacent.
 The number of such environmentally equivalent O sites,  
 which not contribute to the EHC effect, increases with increasing 
 the dopant concentration.

 The high-probability O hopping via $V_0$ between the low-energy 
 equivalent sites, localized close up to each other,
 can be considered as some sort of the vacancy block, which results in 
 the anisotropic absorption peak to the internal frictions 
 measured in 9.5 mol $\%$ Y$_{2}$O$_{3}$.\cite{Ohta+2002} 
 This observation is in agreement with our findings confirming earlier 
 suggestions that some vacancies exist in bound states as
 di-$V_0$ along the [111] fluorite direction.
 As the number of $V_0$-blocks increases with Y$_{2}$O$_{3}$ doping,
 the peak magnitude for the localized relaxation may not scale simply 
 with the dopant concentration, reaching a maximum at some critical 
 concentration $x_m$ correlated with that of the EHC maximum.
 At $x > x_m$, the number of sites on the O sublattice with similar
 local environment increases significantly. This may lead
 to the long-range transport of O ions via vacancies reducing
 therefore the local hopping process in $V_0$-blocks. 
 Regarding the concentration profile of localized relaxation
 explained, such observations at $x > x_m$ may be masked 
 by overlap between the isotropic and anisotropic absorption 
 peaks due to diffusion and $V_0$-block relaxation, respectively. 
           
 In {\it summary}, our study of YSZ being performed using
 the {\it ab initio} relaxed configurations and dynamical model of
 O hopping allows to explain the EHC anomaly observed with
 the Y$_2$O$_3$ doping. In YSZ, as it has been demonstrated, the 
 anisotropic absorption peak of internal frictions and EHC effect 
 have a common vacancy-block nature showing great potential for 
 theory to deal with the realistic model of doped ceramics. 
 Because of the general physics
 involved we believe that similar excessive behaviour of heat capacity
 might be observed in a wide class of ionic conductors 
 where anion hopping goes through non-equivalent positions.   

 E.S. thanks Russian Basic Research Foundation (Grants No.   
 00-02-17426, 01-02-96462) for financial support.


\begin{thebibliography}{99}

\bibitem{Flynn} C.P.~Flynn, {\it Point Defects and Diffusion},
 (Clarendon Press, Oxford, 1972), p. 423.

\bibitem{Subbarao} 
 E.\,C. Subbarao, in {\it Science and Technology 
 of Zirconia}, edited by A.\,H. Heuer and L.\,W. Hobbs, 
 {\it Advances in Ceramics}, Vol. 3 (American Ceramic Society, 
 OH, 1981).

\bibitem{Li} P.~Li {\it et al.}, 
 Phys. Rev. B {\bf 48}, 10063 (1993).

\bibitem{Stefanovich} E.V.~Stefanovich {\it et al.}, 
 Phys. Rev. B {\bf 49}, 11560 (1994). 

\bibitem{OCMVAFP} S.~Ostanin {\it et al.}, 
 Phys. Rev. B {\bf 65}, 224109 (2002).

\bibitem{Ali} A. Alavi {\it et al.}, 
 Phys. Rev. Lett. {\bf 73}, 2599 (1994).

\bibitem{heat_cap} T.~Tojo {\it et al.}, 
 J. Thermal Analysis Calorimetry {\bf 57}, 447 (1999).

\bibitem{Exp-neutron} D.W.~Liu {\it et al.}, 
 Phys. Rev. B {\bf 36}, 9212 (1987);
 D.N.~Argyriou and M.M.~Elcombe,
 J. Phys. Chem. Solids {\bf 57}, 343 (1996).

\bibitem{OS2002} S.~Ostanin {\it et al.}, 
  Phys. Rev. B {\bf 66}, 132105 (2002).

\bibitem{Miller81} R.A.~Miller {\it et al.},
 in {\it Science and Technology of Zirconia},
 edited by A.\,H. Heuer and L.\,W. Hobbs, 
 {\it Advances in Ceramics}, Vol. 3, 
 (American Ceramic Society, OH, 1981).

\bibitem{Bennett} C.H.~Bennett,
 in {\it Diffusion in solids. Recent developments},
 edited by A.S.~Nowick and J.J.~Burton,
 (Academic Press, NY, 1975). 

\bibitem{OCMc+2000} S.~Ostanin {\it et al.}, 
 Phys. Rev. B {\bf 62}, 14728 (2000).

\bibitem{Arrhenius} A.~Cheikh {\it et al.},
 J. Europ. Ceram. Soc. {\bf 21}, 1837 (2001). 

\bibitem{Goff99} J.P.~Goff {\it et al.},
 Phys. Rev. B {\bf 59}, 14202 (1999).

\bibitem{Ohta+2002} M.~Ohta {\it et al.}, 
 Physica B {\bf 316-317}, 427 (2002);
 M.~Ohta {\it et al.},
 Phys. Rev. B {\bf 65}, 174108 (2002).   

\end{thebibliography}
\end{document}